\documentclass[journal=nalefd,manuscript=communication]{achemso}
\usepackage[version=3]{mhchem} 
\usepackage{amsmath}
\usepackage{braket}
\usepackage{array}
\usepackage{graphicx}
\usepackage{caption}
\usepackage{subcaption}
\usepackage{color}
\usepackage{fixltx2e}

\title{Mobile Small Polarons Explain Conductivity in Lithium Titanium Oxide Battery Electrodes}

\author{Matthias Kick}
\affiliation{Chair for Theoretical Chemistry and Catalysis Research Center, Technical University of Munich, Lichtenbergstr. 4, 85747 Garching, Germany}
\author{Cristina Grosu}
\affiliation{Chair for Theoretical Chemistry and Catalysis Research Center, Technical University of Munich, Lichtenbergstr. 4, 85747 Garching, Germany}
\alsoaffiliation{Institute of Energy an Climate Research (IEK-9), Forschungszentrum J\"ulich, 52425 J\"ulich, Germany}
\author{Markus Schuderer}
\author{Christoph Scheurer}
\author{Harald Oberhofer}
\email{harald.oberhofer@tum.de}
\affiliation{Chair for Theoretical Chemistry and Catalysis Research Center, Technical University of Munich, Lichtenbergstr. 4, 85747 Garching, Germany}
\date{\today}

\begin{document}
\begin{abstract}
Lithium titanium oxide Li$_4$Ti$_5$O$_{12}$ (LTO) is an intriguing anode
material promising particularly long lived batteries, due to its remarkable
phase stability during (dis)charging of the cell. However, its usage is
limited by its low intrinsic electronic conductivity. Introducing oxygen
vacancies can be one method to overcome this drawback, possibly by altering
the charge carrier transport mechanism.  We use Hubbard corrected
density-functional theory (DFT+U) to show that polaronic states in combination
with a possible hopping mechanism can play a crucial role in the
experimentally observed  increase of electronic conductivity. To gauge
polaronic charge mobility, we compute relative stabilities of different
localization patterns and estimate polaron hopping barrier heights.
With this we finally show how defect engineering can indeed raise the 
electronic conductivity of LTO up to the level of its ionic conductivity,
thereby explaining first experimental results for reduced LTO.
\end{abstract}

\section{Introduction}
Energy storage solutions such as Li-ion batteries (LIB) are a key technology
in the transition from a fossil fuel based economy to a society based on
sustainable resource management.\cite{C3CS60199D,ZHAO20151} 
Despite the tremendous advancements in battery research over the last few
years,\cite{Chu2012}
durability and especially storage capacity still need significant improvements
for batteries to represent a viable alternative e.g.~in the transport and
mobility sectors.\cite{Armand2008,ZHAO20151}
One promising material envisioned as a potential remedy for these problems in
conventional as well as all-solid state batteries (ASSB) is lithium titanium
oxide (LTO).\cite{ZHAO20151}
Zero strain insertion, high cycling stability and a stable charge/discharge
plateau render LTO an excellent anode material for long living
batteries.\cite{Armand2008,Arico2005,hendrik1,ZHAO20151}
Its general use, however, is still limited by the fact that LTO suffers from a very low intrinsic
electronic conductivity.\cite{YUAN20104997,C5TA00887E}
One way to overcome this drawback is to expose LTO to a reductive hydrogen
atmosphere at elevated temperatures, leading to the formation of oxygen
vacancies. As experimental data shows \cite{blue_LTO}, this not only causes a
color change from white to blue but also lowers electronic resistance and
impedance. Moreover, this blue LTO also shows improved Li-ion mobility
compared to pristine white LTO.\cite{blue_LTO}

Unfortunately, neither of these improvements in carrier mobility are currently
fully understood from a mechanistic viewpoint.
Yet, first hints at the nature of the improved electronic conductivity in LTO
emerged recently with the experimental discovery of paramagnetic
Ti\textsuperscript{3+} centers.\cite{blue_LTO,C5TA00887E,C3TA11590A,Peter2015}
The significance of these becomes apparent considering an analogous case in
TiO\textsubscript{2}, where oxygen vacancies are known to lead to the
formation of small polarons mainly localized on Ti
sites.\cite{dupuis,0953-8984-24-43-435504,PhysRevLett.113.086402}
While these polarons are somewhat attracted to the vacancy
itself,\cite{Kick2019JCTC,PhysRevB.98.045306}
they were also shown to be very mobile, with kinetic barriers that can easily
be overcome at room temperature.\cite{PhysRevLett.105.146405}
In this context we studied the formation and stability of polarons in bulk
LTO, as well as the kinetic barriers separating them.  Especially the latter
strongly hints at a polaron hopping mechanism as the source of the observed
improvement of electronic conductivity in blue LTO.

\section{Results}
\begin{figure}
\includegraphics[scale=.25]{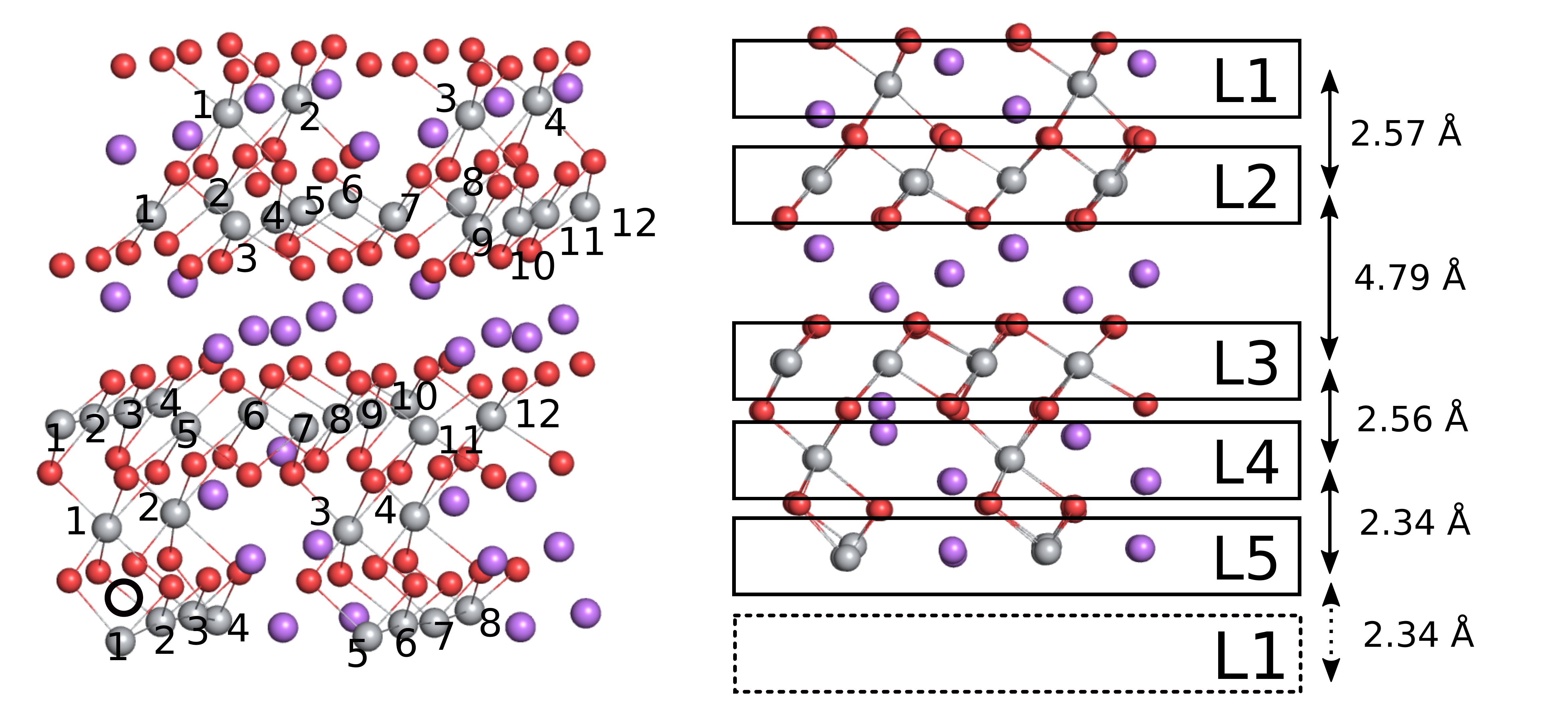}
\caption{Sketch of the site naming convention of the different localization patterns. L$x$-$m$/L$y$-$n$ specifies the localization of one electron within layer L$x$ on atom $m$ with the second electron localized in layer L$y$ on atom $n$. Additionally, the position of the oxygen vacancies is marked with a black circle. Titanium atoms are depicted as grey spheres, oxygen atoms are shown as red spheres. On the right, arrows and numbers indicate the average inter layer titanium distance within the given unit cell.}
\label{site_naming}
\end{figure}

For our analysis we considered a $2\times 2\times 1$ supercell of Li$_4$Ti$_5$O$_{12}$ 
(LTO) in its most stable\citep{hendrik1} spinel configuration,\citep{DESCHANVRES1971699} 
and created an oxygen vacancy at the energetically most favorable
of the symmetry inequivalent sites (cf.~supplementary material).
This realization of the structural Li/Ti disorder is known to exhibit a Ti-deficient zone separating titanium layers as
illustrated in Figure~\ref{site_naming}. It therefore lets us examine the interplay
between structure, in the form of Li-rich and Li-poor regions, and function of the material.
Removing a neutral oxygen atom from the simulation cell gives rise to two
excess electrons, which can form two polaronic Ti\textsuperscript{3+} centers.
In standard semi-local LDA\citep{hohenberg_kohn} or GGA\citep{PBE} based DFT
these can generally not be described at all due to the functionals' well known
charge and spin delocalization errors.\citep{Cohen2008SCI} As a cost-efficient
remedy, we here make use of the popular Hubbard corrected
\citep{Hubbard238,anisimov_1,dudarev} variant of the PBE\citep{PBE} functional (PBE+U) in
combination with the matrix control
approach.\citep{PhysRevB.79.235125,matrix_control}
This combination has not only been shown to yield easy access to all manners
of polaron configurations but also to yield excellent results compared to the
computationally much more expensive hybrid DFT functionals.\citep{Kick2019JCTC}
Given the great structural complexity of defect-rich LTO, there is a large
number---$\binom{40}{2}=780$---of unique polaron localization patterns even in
our relatively small simulation box. To distinguish them we calculated their
relative stability according to
$
E_{\rm rel} = E_{\rm tot, i } - E_{\rm tot,\;min} \quad,
$
where $E_{\rm tot, i }$ denotes the total energy of a given simulation box
calculated with DFT and $E_{\rm tot,\;min}$ denotes the total energy of
the most stable structure found so far. Hence, following standard procedures,\citep{dupuis}
the most stable configuration serves as zero point of our energy scale, with
all other configurations possessing positive relative energy.

\begin{figure}
\begin{tabular}{cc} 
\includegraphics[scale=.18]{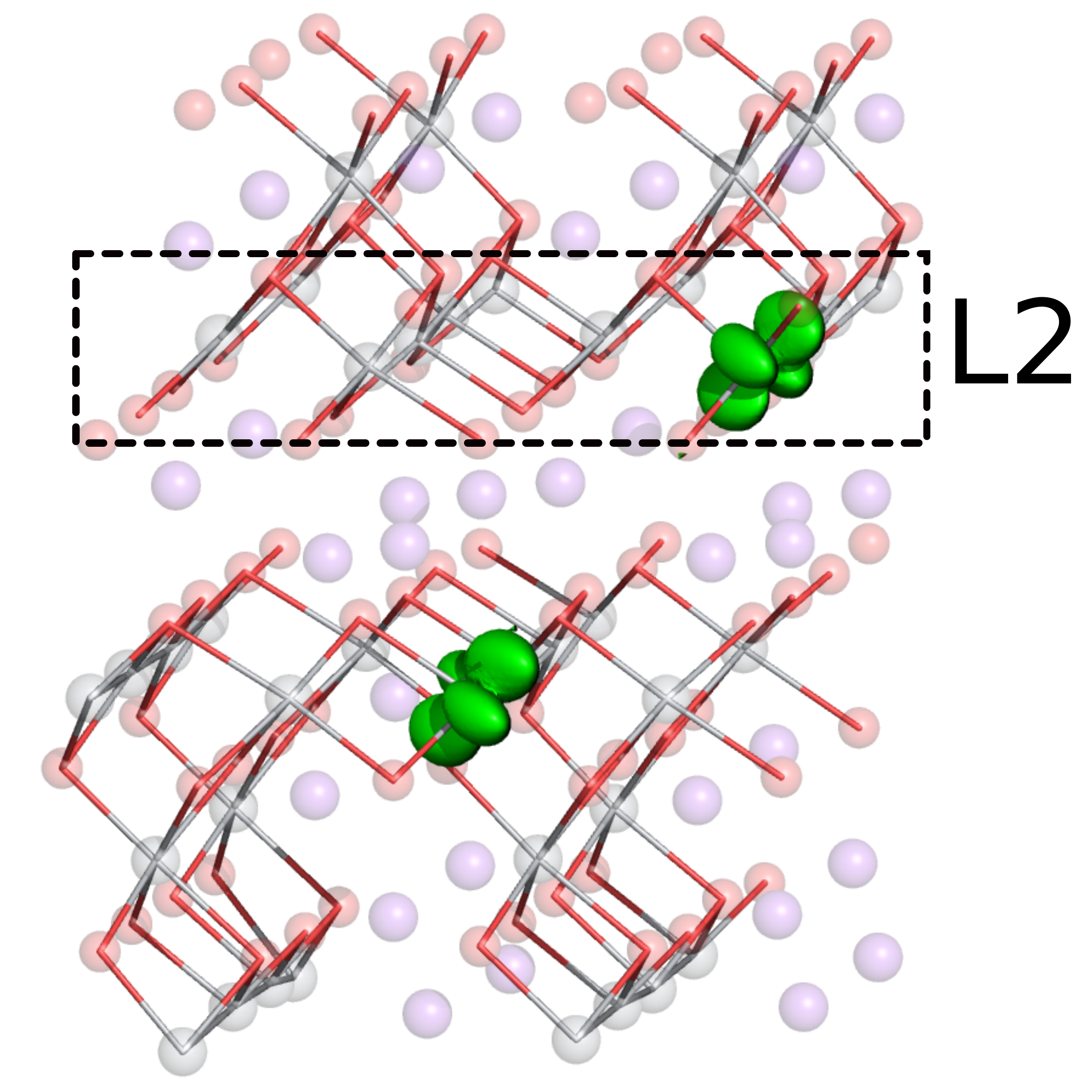} & \includegraphics[scale=.4]{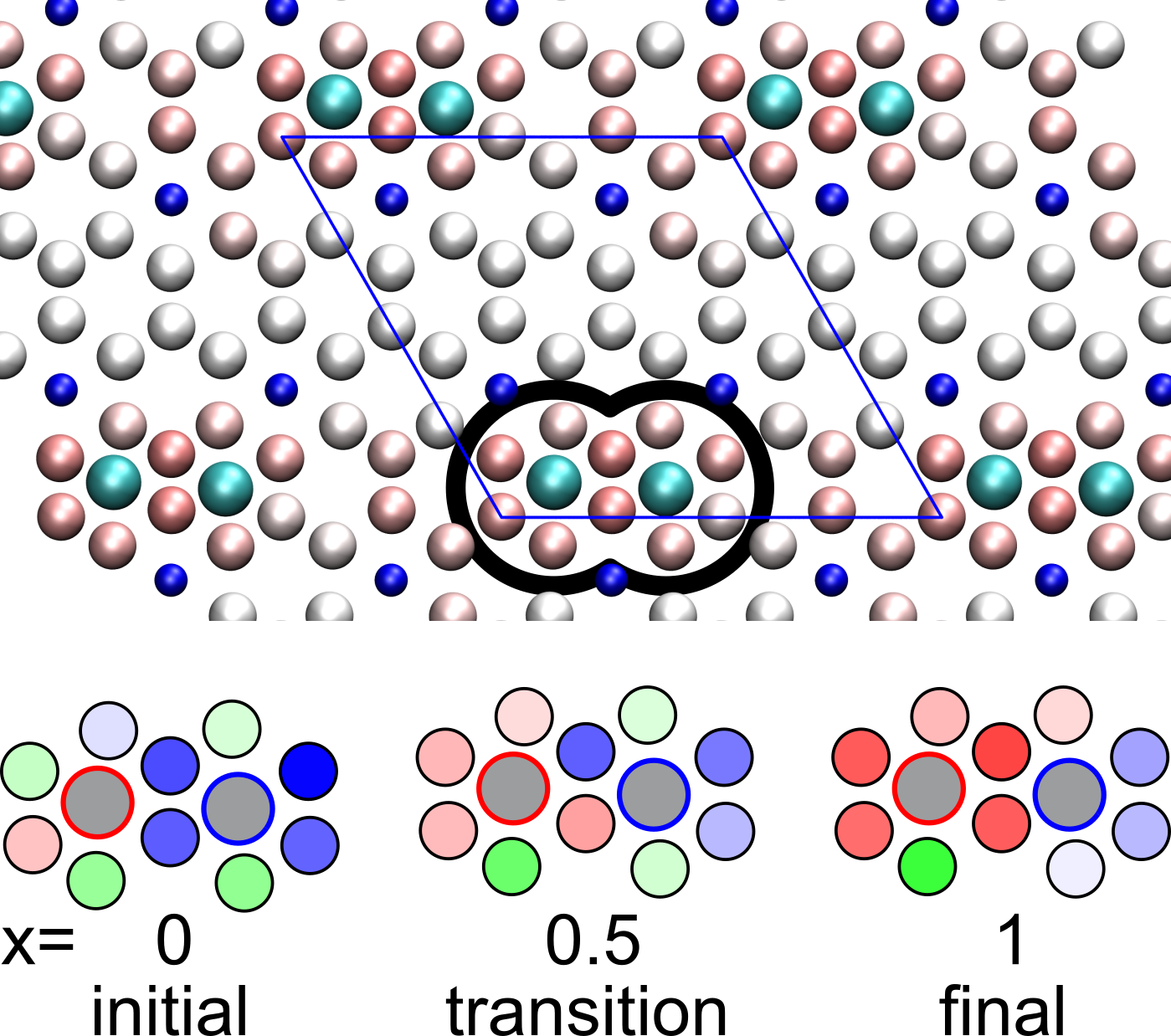}  \\ 
a) & b) \\ 
\end{tabular}
\caption{a) Spin density of the most stable configuration L3-7/L2-9 of our simulation. Isosurface level 0.015 e$\rm\AA^{-3}$. b) Top: cut through the simulation cell showing the L2-layer, showing the distortion of the lattice at the transition state. Atoms in cyan indicating the L2-9 and L2-10 positions respectively. Darker red colors indicate a stronger movement during the transition from L3-7/L2-10 to L3-7/L2-9. Note that only nearest neighbor oxygen atoms show significant movement, indicating indeed the hopping of a small polaron. Bottom: schematic of the movement of oxygen atoms from their undistorted sites for the transition from L3-7/L2-10 (left) to L3-7/L2-9 (right) via a transition state (center). Circles filled in red hues indicate a predominant movement of the respective O atom towards the Ti atom at L2-9 (red circle filled with grey), while blue hues depict a movement towards L2-10 (blue circle filled with grey). Circles filled in green hues show movement not clearly aimed at either Ti atoms. In all cases darker colors indicate stronger movement.}
\label{most_stable_configuration}
\end{figure}

An exhaustive computational sampling is complicated further by the fact that the electrons can
localize both in a triplet or an open-shell singlet
configuration.\citep{dupuis,0953-8984-24-43-435504} On the other hand, many of
these patterns are, if not fully degenerate, at least very close in energy. For a first
demonstration of the existence and mobility of polarons in LTO and their influence on the
electronic conductivity a complete
sampling of all configurations is therefore not necessary. Instead, we focus
on triplet configurations and localization patterns representative for the
system as a whole. In detail, we considered 13 patterns with different
distances to the defect site, localized within different Ti layers in the bulk
unit cell. To distinguish them, we adapt a naming convention used in our
earlier work\citep{Kick2019JCTC} to the case of LTO, cf.~Figure~\ref{site_naming}.
We found the most stable defect position to be located in the center of the Ti-rich
region of our layered LTO model (black circle in Figure~\ref{site_naming}, cf.~also supplementary material).
This localization is not at all surprising, as it allows the O-vacancy to be as
far as possible from the Li-rich zone of our simulation cell and the structural distortions caused by it.
Using this defect position throughout, Table \ref{rel_stabilities} lists our obtained results regarding the relative stability of different polaron localization patterns.
\begin{table}
\caption{Relative stabilities of the most representative polaronic configurations.
The systems are ordered by their relative stabilities.
A complete list of calculated systems can be found in the supplementary
information. Also shown is the shortest periodic distance between the two
Ti\textsuperscript{3+} centers (d$_{\rm Ti^{3+}-Ti^{3+}}$).} For vertical
layer--layer distances see Fig.~\ref{site_naming}.
\label{rel_stabilities}
\begin{tabular}{lccclcc}
\hline\hline
system &  E$_{\rm rel}$ [eV] & d$_{\rm Ti^{3+}-Ti^{3+}}$[\AA] &  \rule{5mm}{0pt} & system &  E$_{\rm rel}$ [eV] & d$_{\rm Ti^{3+}-Ti^{3+}}$[\AA] \\ 
\hline 
L3-7/L2-9  & 0.00 & 6.6 & & L5-1/L3-12 & 0.45 & 7.9\\
L3-8/L3-4  & 0.12 & 6.0 & & L4-2/L4-4  & 0.75 & 5.9\\
L3-7/L3-12 & 0.13 & 7.9 & & L1-1/L1-4  & 0.77 & 10.0\\
L2-7/L2-8  & 0.14 & 5.9 & & L5-5/L4-2  & 0.82 & 7.9\\
L3-9/L3-12 & 0.23 & 3.0 & & L4-3/L4-4  & 1.00 & 6.0\\
L2-9/L2-12 & 0.26 & 2.9 & & L5-1/L5-5  & 2.59 & 6.0\\
\hline\hline
\end{tabular} 
\end{table}
We found that the most stable configurations are those where one polaron is
located in L2 and the other is located in L3 (cf.~Figure~\ref{most_stable_configuration}a),
followed by configurations where both polarons are either located in L2 or
L3. The main difference is that ``same-plane'' polarons approach each other
more closely and hence Coulomb repulsion is more pronounced compared to the
most stable L3-X/L2-Y configurations. Furthermore, our analysis shows that
there is also a tendency for polarons to be less stable the closer they
are located to the defect site. This is clearly indicated by the configurations
where both excess electrons are localized on Ti atoms belonging to layers L4
or L5 (see supplementary information for detailed distances to the defect site). 
Both of these effects compete with each other, such that polarons
try to adopt configurations with maximal distances to the defect and between each other.
This situation seems similar to TiO\textsubscript{2} where the oxygen defect is 
acting as a charge trapping center.\cite{Kohtani_2017,Diebold200353} 
However, in LTO one can not directly extract a clear stability trend with 
the distance to the defect, as positively charged Li-ions also show 
some influence on the overall stabilities of the polarons.\citep{PhysRevB.82.144119}
Indeed, our choice of LTO cell allows us to quantify their influence, considering
the fact that all our most stable defect configurations are located next to the Li-rich 
zone situated between L2 and L3.

Finally, in order to gauge the polaron mobility we calculate hopping barriers
between our most stable configuration and adjacent Ti atoms. To this end, we
again make use of the matrix control approach,\citep{PhysRevB.79.235125,matrix_control}
but with a modified occupation matrix scheme outlined in the supplementary material.
This approach allows us to restrain the electronic configuration of the system
along a pre-selected reaction coordinate x, which linearly interpolates between two stable polaronic 
states, localized at neighboring atoms.
Representatively for all hopping processes in the system we compute the ``in-plane''
transition of L3-7/L2-10 (x = 0.0) to L3-7/L2-9 (x = 1.0) and the ``out-of-plane'' transition
of  L3-7/L1-3 (x = 0.0) to L3-7/L2-9 (x = 1.0). 

We illustrate this pathway in Figure~\ref{most_stable_configuration}, which in a) shows the spin
density of the final state L3-7/L2-9 and in b) depicts we the relaxation of atoms from their respective
sites in the pristine crystal for the transition from L3-7/L2-10 to L3-7/L2-9 via a transition state. 
We thereby only depict O atoms nearest to the involved Ti sites as only these show any significant 
distortion during a full geometry optimization. 
Figure~\ref{most_stable_configuration}b clearly shows relaxation of the O-atoms
towards the respective Ti$^{3+}$ sites in the initial (L2-10, blue) and final (L2-9, red) states, 
indicating a small polaron hopping mechanism.
A similar picture arises for the transition from L3-7/L1-3 to L3-7/L2-9 (not pictured).

\begin{figure}
\includegraphics[scale=.6]{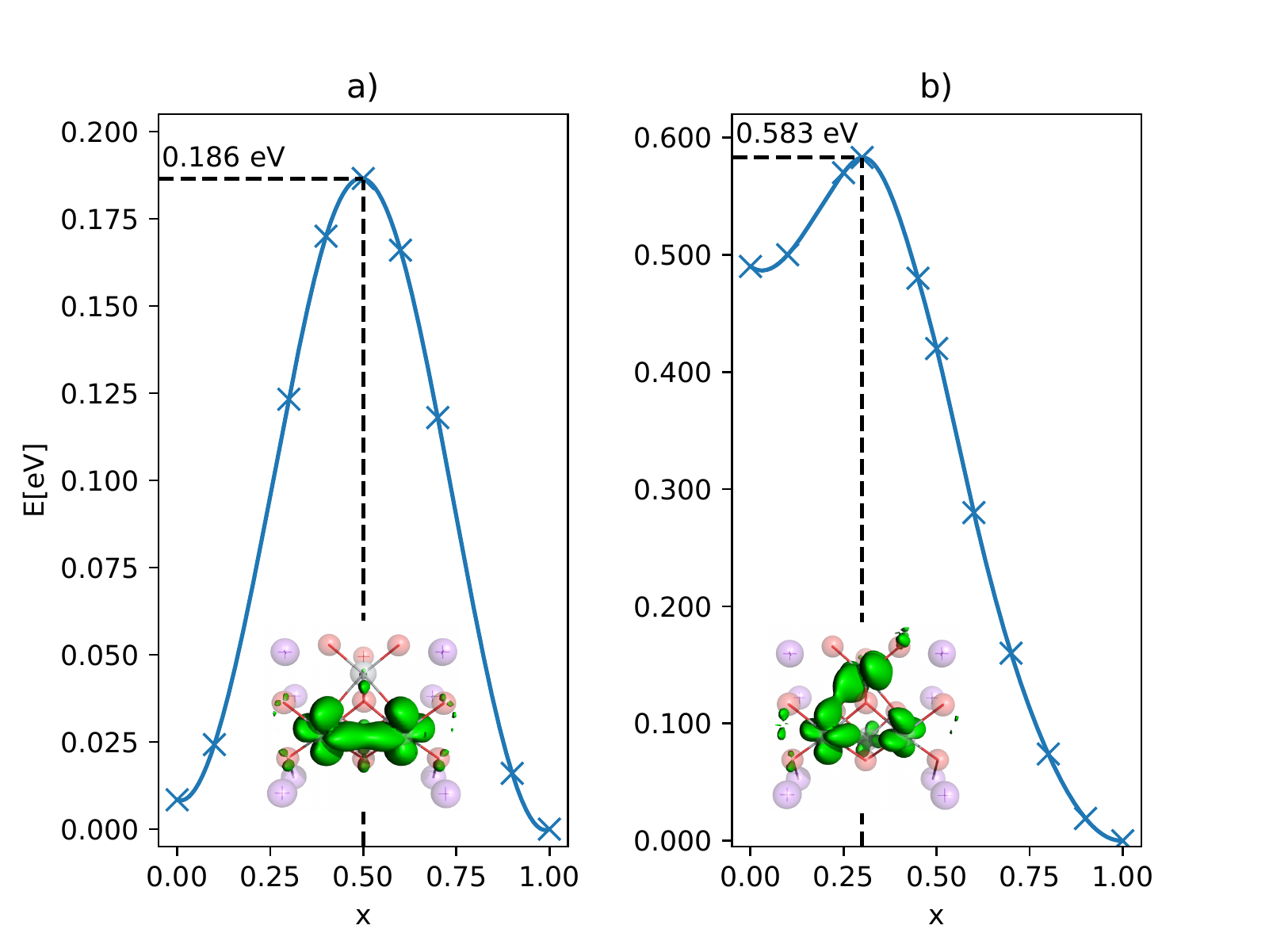}
\caption{Calculated barrier profiles for the transition L3-7/L2-10 to L3-7/L2-9 (a) and L3-7/L1-3 to L3-7/L2-9 (b). In both plots x = 1.0 is equal to configuration L3-7/L2-9.  The lowest lying configuration in energy serves as zero point for the energy scale. Also shown is the spin density of the corresponding transition state. Isosurface level 0.015 e$\rm\AA^{-3}$.}
\label{barrier_profile}
\end{figure}

The energy profiles of these two transitions are depicted in Figure \ref{barrier_profile}. 
As highlighted by the dashed black lines, the hopping barriers for transition L3-7/L2-10 to L3-7/L2-9 
and for transition L3-7/L1-3 to L3-7/L2-9 are 186\,meV and 583\,meV respectively, with the later 
being much larger due to the already significant energy difference of 485\,meV between the two stable states. 
With these barrier heights we can roughly estimate the in- and out-of-plane conductivity based on a
simple hopping model (cf.~supplementary material) and using an experimentally measured density of 
Ti\textsuperscript{3+} centers of $13.1\,\text{at}\%$\citep{nasara} as a measure for the charge carrier density. For the in-plane conductivity we thus find a value of $95.3$ mS/cm, while the significantly higher out-of-plane hopping barrier results in a 
much smaller conductivity of $17\times 10^{-6}$ mS/cm. 
To put these results into context, even our lower bound for the conductivity of reduced LTO is already five orders of magnitude higher than the pristine material,\citep{nasara} while our ideal upper bound is of the order
of the ion conductivities in currently employed electrolytes.\cite{PARK20107904}
Note that the estimate for the ideal conductivity rests on the assumption that there are no other, significantly higher barriers along the whole pathway of charge percolation through the crystal. This implies a distribution of defects aligned along the [100] axis of the crystal.
Considering the fairly high density of oxygen defects present in blue LTO, such a case is certainly achievable.

\section{Discussion}
To conclude, our stability analysis clearly indicates that the experimentally observed Ti\textsuperscript{3+} 
centers in reduced LTO can in fact be the result of small polaron formation. Moreover, comparatively small 
barrier heights indicate that charge hopping dynamics can already occur at room temperature. 
This renders a polaron hopping mechanism to be the most likely origin of the increased 
electronic conductivity observed in blue LTO. Indeed a simple conductivity model puts even the worst
estimate of $17\times 10^{-6}$ mS/cm five orders of magnitude above pristine LTO.
On the other hand, the ideal predicted case based on an in-plane hopping mechanism would lead to a conductivity 
of $100$ mS/cm, which, though significantly lower than that of other anode materials
\cite{PhysRev.90.187,PARK20107904,DT9740002325}, is nevertheless comparable to the ion conductivities 
of the pristine material and super--ionic conductors.\cite{Stenina2015,C3TA11590A,ZHAO20151} Moreover, the existence of polarons
also hints at a mechanism for the improved ion diffusivity in blue LTO, which is about twice that of pristine LTO.\citep{nasara} 
The presence of polarons, which we have shown to localize near Li-rich regions, could serve to ``soften'' the environment for Li-diffusion by screening the positive charge carriers.
Thus, both, the ideal polaron conductivity of blue LTO, and its improved Li diffusivity, would make it 
a suitable option for an anode material for use in tomorrow's batteries. Note that our 
results show a very wide range of potential conductivity values, depending on defect patterning, local crystal
structure and crystal orientation. Our study thus highlights the potential and also pitfalls of defect 
engineering as a means for the generation of mobile charge carriers in otherwise insulating materials.

%
\section{Methods}
All necessary calculations have been carried out using the FHI-aims code.\citep{Blum20092175} To achieve adequate 
electron localization we used the DFT+U\cite{anisimov_1} variant of the PBE \citep{PBE} exchange correlation 
functional. In detail we used the rotationally invariant +U  form \citep{dudarev} with the double-counting correction in 
the fully localized limit.\citep{QUA:QUA24521}
The Ti $3d$ atomic like basis functions served as Hubbard projectors and a U value of 2.65 eV has been applied. 
Numerical convergence has been reached using a \textit{tight} \textit{tier1} basis---roughly equivalent to a polarized 
triple zeta split valence Gaussian type orbital basis set\citep{Lamiel2017JCTC}---employing a 2 $\times$ 2 $\times$ 
1 k-point grid. Geometries have been relaxed until residual force fell below $10^{-2}$ eV/\AA{}. A detailed 
description of the methodology used to estimate barriers is given in the supplementary material.

\section{Acknowledgements}
The authors would like to thank the German Research Foundation DFG (Grant OB425/4-1) and the Solar Technologies Go Hybrid Initiative of the State of Bavaria for support.
Partially funded by the Deutsche Forschungsgemeinschaft (DFG, German Research Foundation)
under Germany's Excellence Strategy – EXC 2089/1 – 390776260.
\bibliography{bib.bib}
\end{document}


%
%
\section{Structures}
\subsection{Pristine bulk structures}
Spinel lithium titanium oxide (LTO, Li\textsubscript{4}Ti\textsubscript{8}O\textsubscript{12}) crystallizes in the Fd3$\rm \overline{m}$  (No. 227) space group. The O atoms form a face center cubic packing (fcc) and occupy the 32e sites within the cubic unit cell. The titanium atoms and one quarter of the lithium atoms are octahedral cooordinated by oxygen and occupy the 16d sites. The ratio between Li and Ti atoms at these octahedral sites is  $\frac{1}{6}$:$\frac{5}{6}$. The remaining Li atoms occupy the 8a sites.\citep{Scharner01031999}

\begin{figure}
\begin{tabular}{ccc}

\includegraphics[scale=.3]{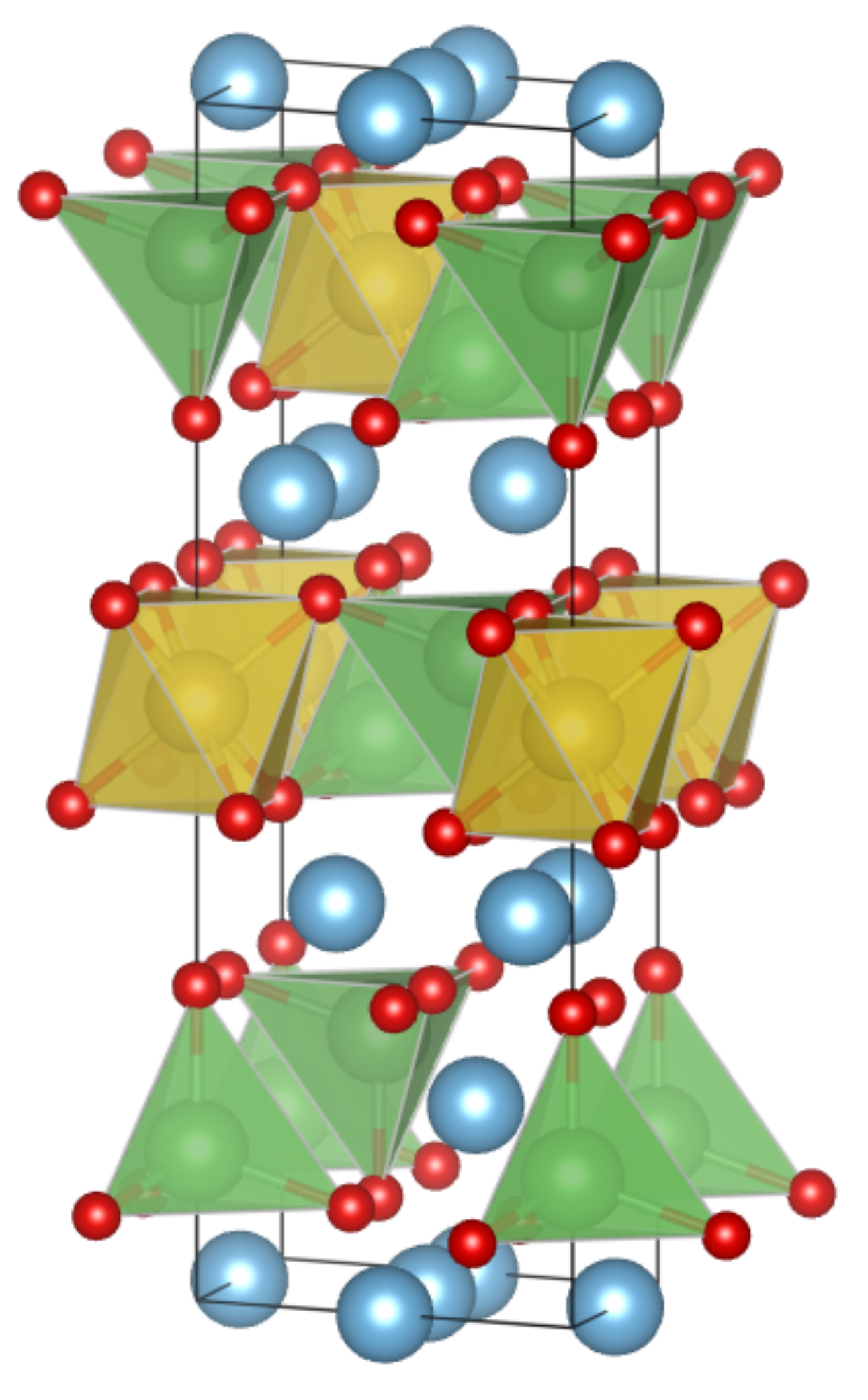} & \includegraphics[scale=.3]{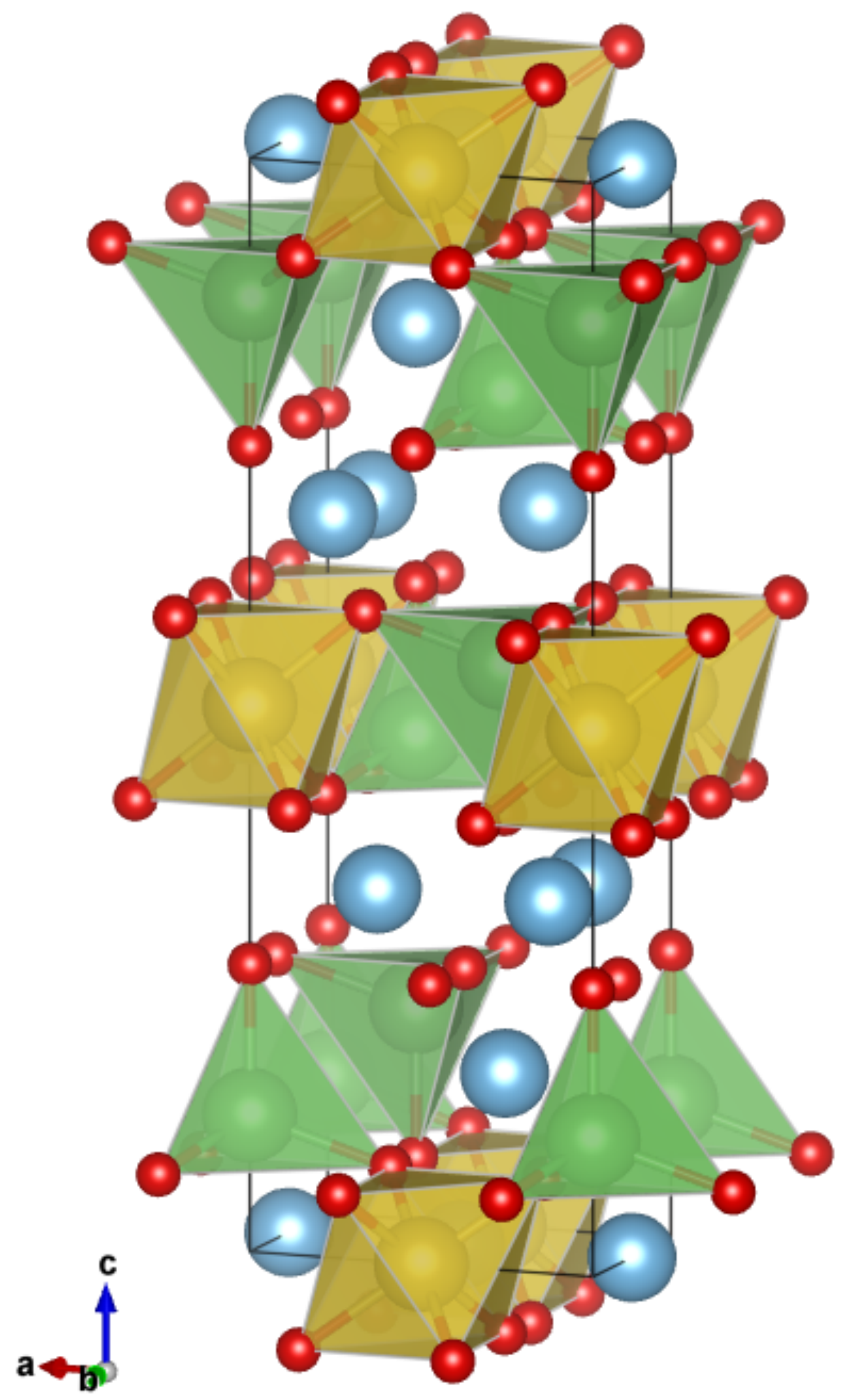} & \includegraphics[scale=.3]{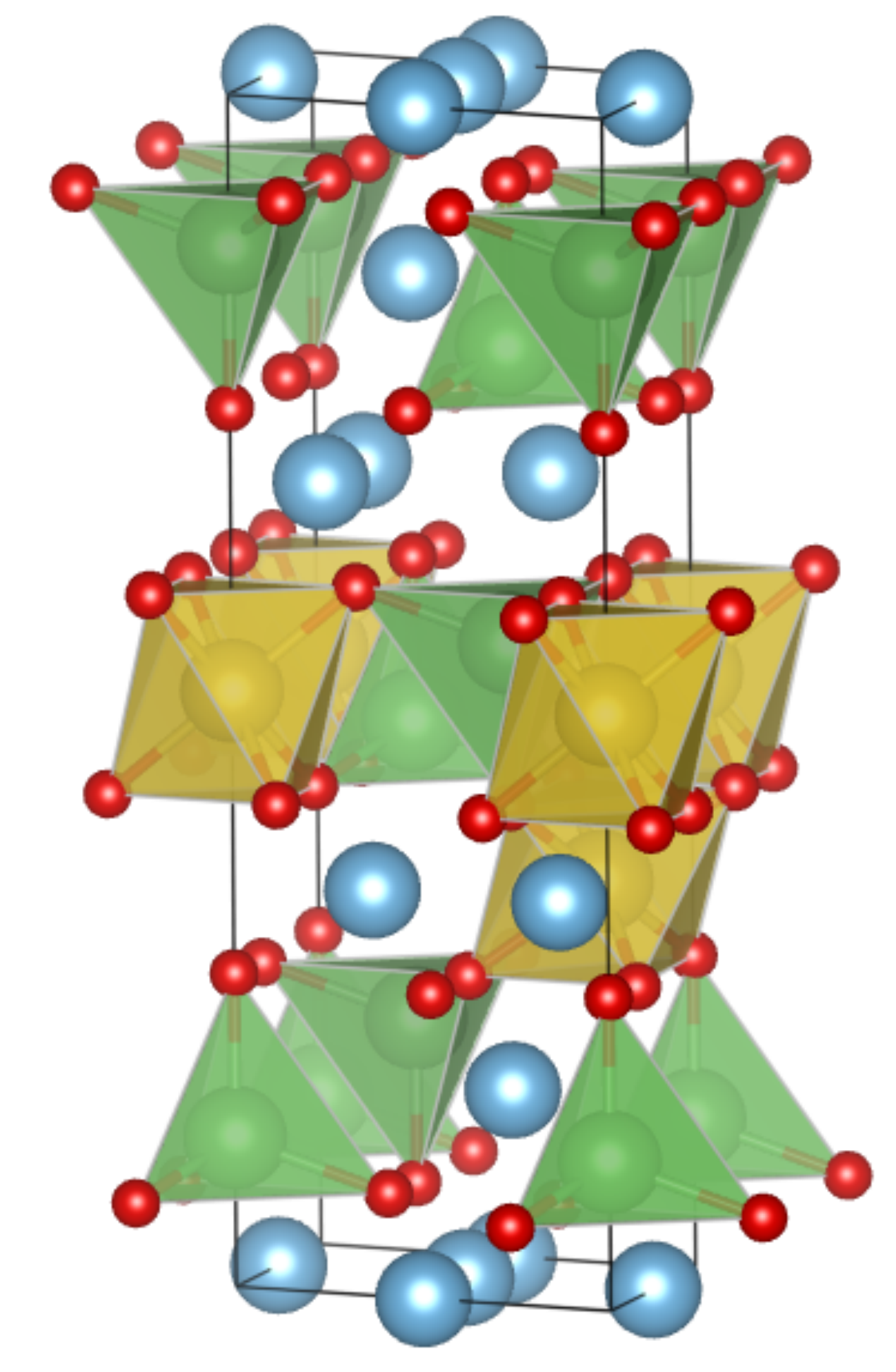} \\ 
c1 & c2 & c3 \\ 
\includegraphics[scale=.3]{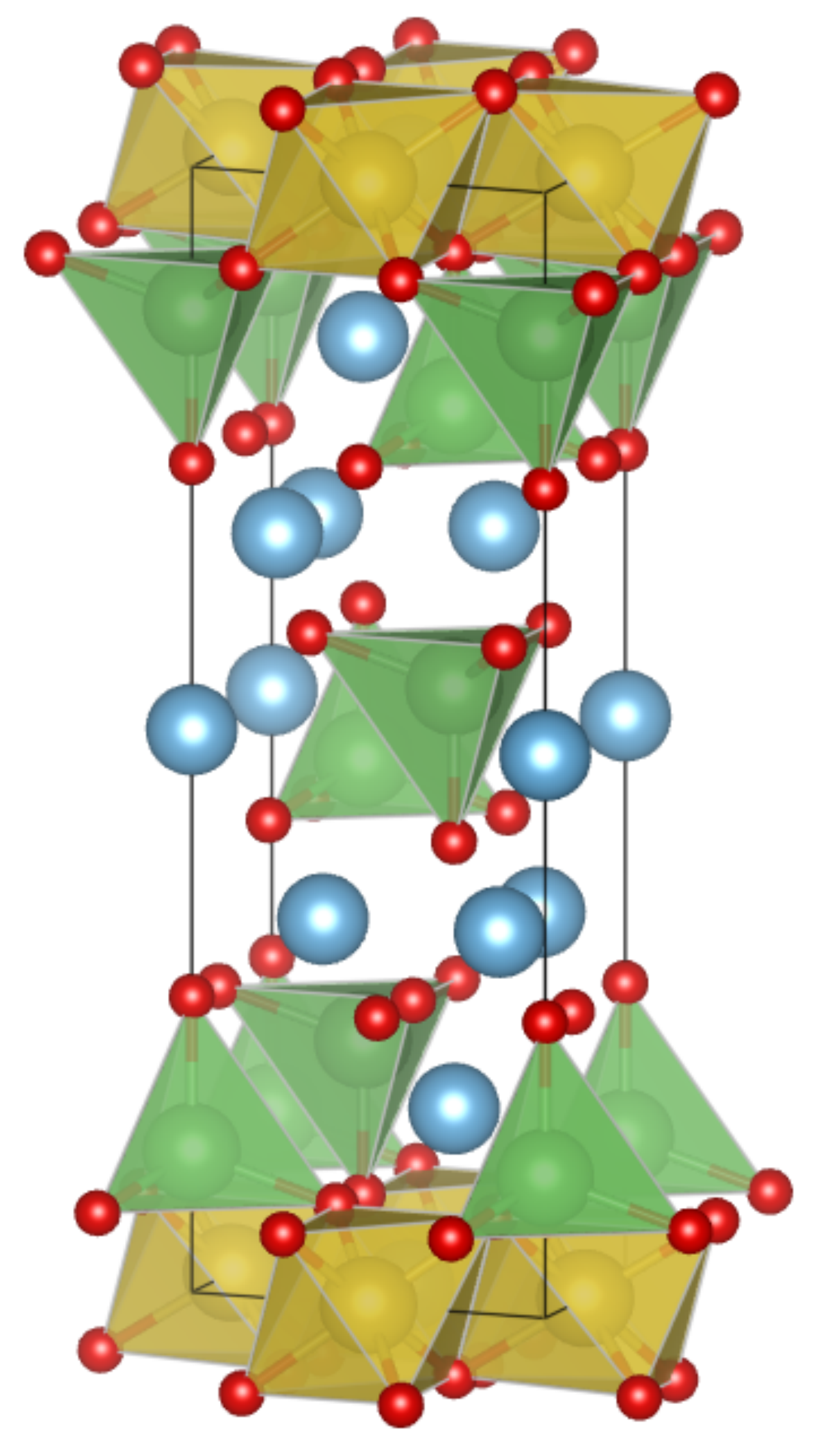} & \includegraphics[scale=.3]{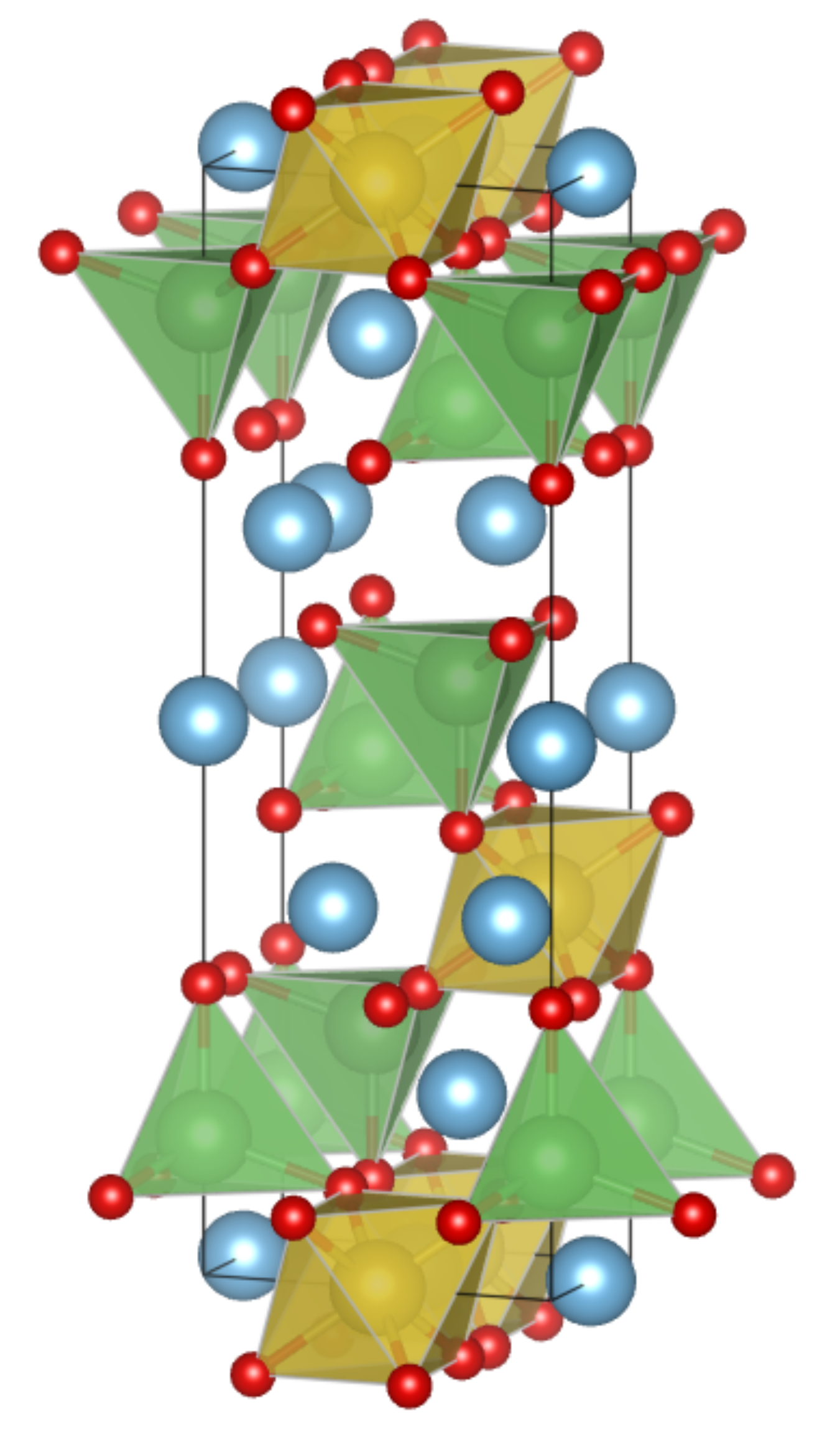} & \includegraphics[scale=.3]{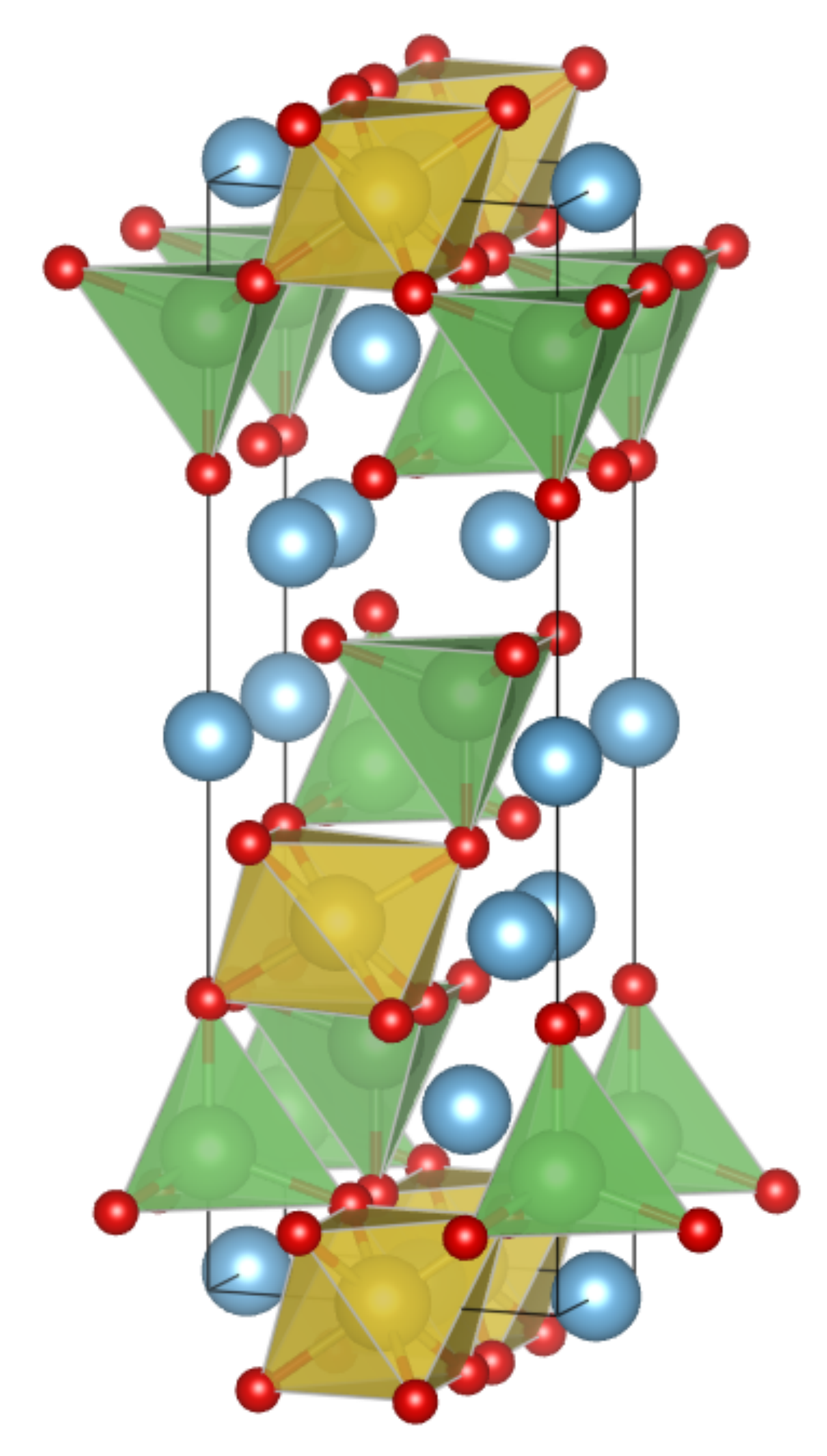} \\ 
c4 & c5 & c6 \\
\end{tabular} 
\caption{Shown are the different LTO unit cells. Ti atoms are shown in light blue, oxygen atoms are red and Li atoms are shown as green spheres. Tetrahedral and octahedral sites are depicted accordingly.}
\label{LTOc_structure}
\end{figure}

The mixed occupancy of Ti and Li at the 16d sites can not be realized with the conventional cubic unit cell of Fd3$\rm\overline{m}$ as there are only sixteen possible sites resulting in a no integer ratio of Li and Ti atoms if the mixed occupancy would be considered.
A suitable subgroup that allows for the right stoichiometry within a single unit cell is R3$\rm \overline{m}$ (No. 166). The unit cell of this space group is hexagonal. Within this cell the mixed occupancy can be easily realized. It contains 12 sites corresponding to the 16d sites in Fd3$\rm \overline{m}$. Among them one has to distribute two Li atoms and ten Ti atoms to achieve the correct ratio. In total this yields $\binom{12}{2}$ = 66 possibilities. Among these one can identify 6 unique structures the others are equivalent by symmetry. All of those are different in energy and are denoted c0001 to c0006. All LTO structure are shown in Figure \ref{LTOc_structure}.

These configurations are the starting point for our analysis. We performed pure PBE\citep{PBE} calculations. Table \ref{rel_structural_stabilities} list the obtained results. Among all systems, c0002 is identified to be the most stable configuration. Therefore, we continue our analysis using the c0002 structure.
\begin{table}
\caption{Relative stabilities, E$^{\rm LTO}_{\rm rel}$, of the different pristine bulk structures. Most stable structure c0002 serves as zero point for the energy scale.}
\label{rel_structural_stabilities}
\begin{tabular}{c c}
\hline\hline 
system & E$^{\rm LTO}_{\rm rel}$[eV] \\ 
c0001 & 0.72 \\
c0002 & 0.00 \\
c0003 & 2.14 \\
c0004 & 4.38 \\
c0005 & 0.77 \\
c0006 & 0.36 \\ 
\hline \hline
\end{tabular} 
\end{table}
\subsubsection{Computational details}
Calculations have been performed using the PBE\citep{PBE} functional as it is implemented in the CASTEP\citep{CASTEP} code using a 4 x 4 x 2 k-point grid. Structures have been fully relaxed until forces where below 10\textsuperscript{-2} eV/\AA \ in combination with a plane wave basis set using an energy cut-off of 600 eV.
\subsection{Defect bulk structures}
\label{bulk_defect_structures}

\begin{figure}
\includegraphics[scale=.35]{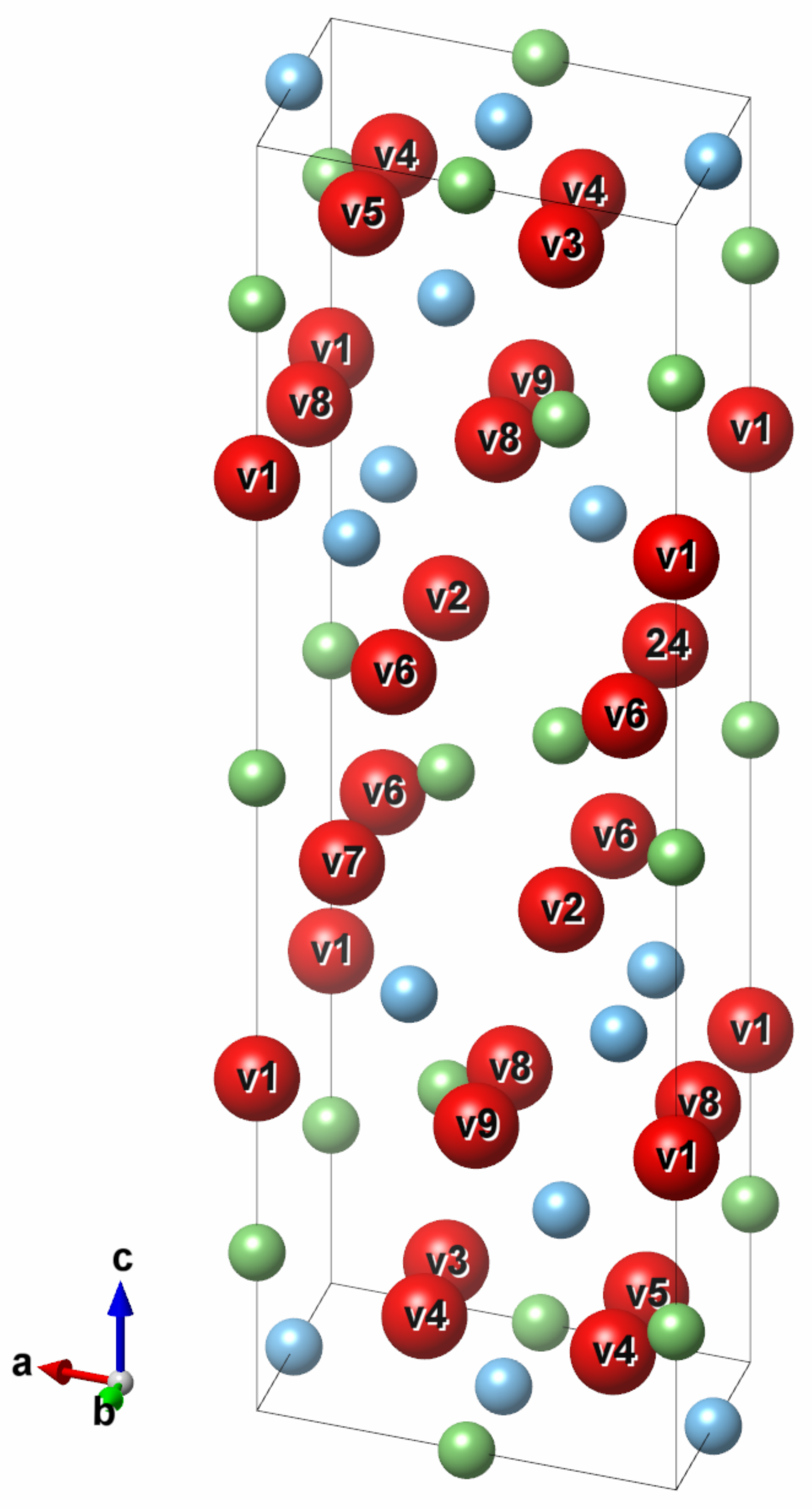}
\caption{Single unit cell of the c2 LTO structure. Ti atoms are shown in light blue, Li atoms are shown in green and oxygen atoms are displayed as red spheres. The different position for an oxygen vacancy are named from v1 to v9. Double naming indicates a symmetry equivalent site.}
\label{defect_sites}
\end{figure}

Within the c0002 unit cell one can identify nine unique O vacancies all others are equivalent due to symmetry (see Figure~\ref{defect_sites}). For this purpose the Site Occupancy Disorder (SOD) package \cite{Grau_Crespo_2007} has been used.

We embedded each of the nine unique oxygen vacancies into 2 x 2 x 1 supercells, yielding a supercell based on three standard defect free unit cells and one with O vacancy.
This structural variety of LTO makes computational sampling rather demanding, therefore, as a first starting point we performed full geometry relaxations using only the PBE functional within FHI-aims. Our results are showed in table \ref{rel_structural_stabilities_LTO_defects} where the different structures are named from v1 to v9 according to the position of the oxygen vacancy in the primitive cell. Again, the most stable configuration serves as zero point for the energy scale.

We like to highlight, that in principle one has to consider here all possible configurations of this system to find the true configuration which is lowest in energy. Therefore, this should be seen just as a first attempt to tackle the problem. Considering, neutral, single and doubly charged oxygen defects in combination with all possible localization patterns of polarons is by far not tractable as one has to go beyond LDA or GGA approximations in order to correctly describe the defect states.

\begin{figure}
\includegraphics[scale=.45]{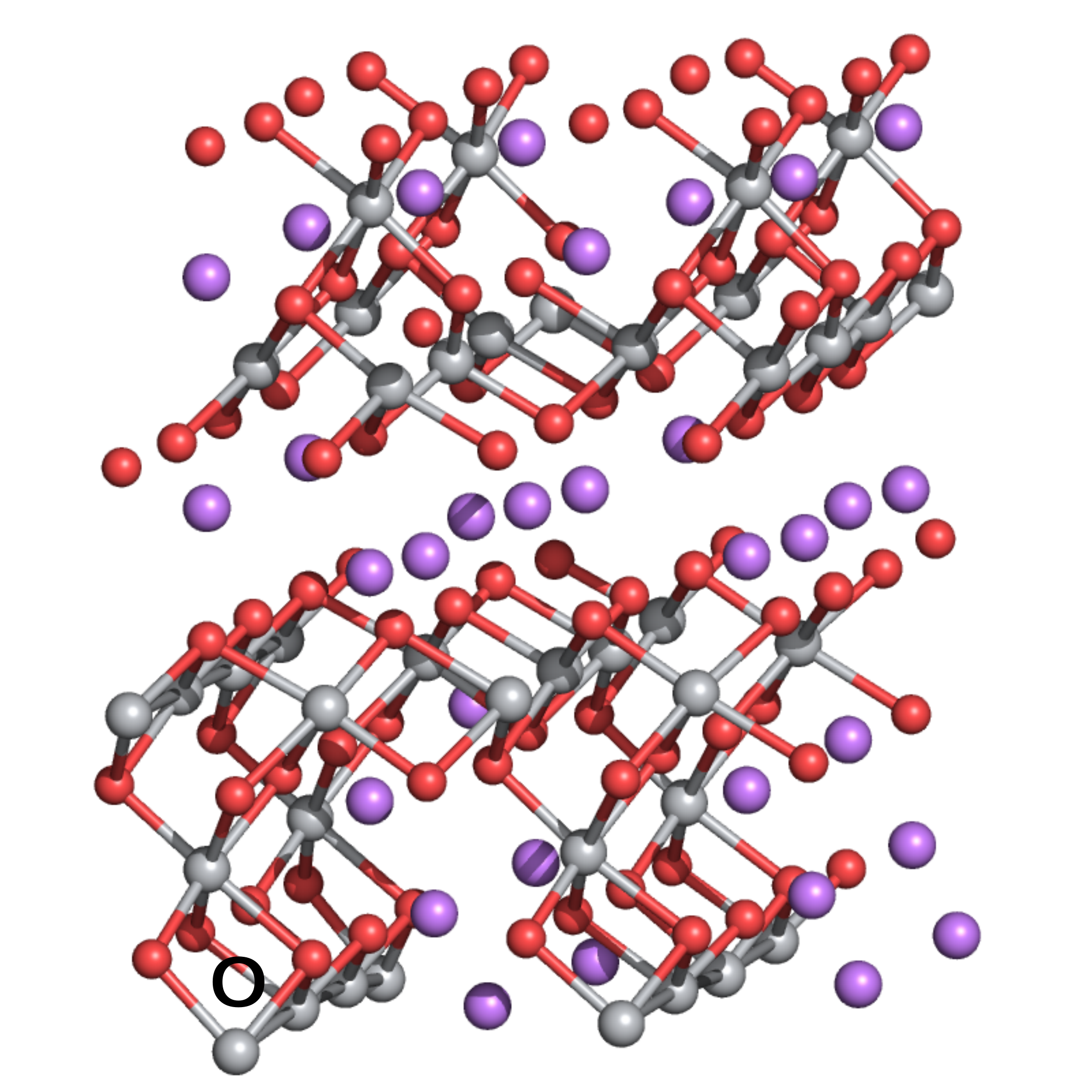}
\caption{Supercell structure consisting of four LTO c2 cells including the v3 vacancy. The O vacancy is marked with a black circle. Li atoms are purple. O atoms are shown as red spheres. Ti atoms are shown as grey spheres.}
\label{v3_structure}
\end{figure}

Among all nine different vacancy configurations we use the most stable one (v3) as the starting point for all further polaron calculations. The supercell of configuration v3 is shown in Figure \ref{v3_structure}.

\begin{table}
\caption{Relative stabilities, E$^{\rm LTO+O_v}_{\rm rel}$, of the different bulk structures with O vacancy. Most stable structure v3 serves as zero point for the energy scale.}
\label{rel_structural_stabilities_LTO_defects}
\begin{tabular}{c c}
\hline\hline 
system & E$^{\rm LTO+O_v}_{\rm rel}$[eV] \\ 
v1 & 1.58 \\
v2 & 1.95 \\
v3 & 0.00 \\
v4 & 0.38 \\
v5 & 1.06 \\
v6 & 0.87 \\ 
v7 & 1.13 \\
v8 & 0.78 \\
v9 & 1.13 \\
\hline \hline
\end{tabular} 
\end{table}
\subsubsection{Computational details}
All calculations have been performed using the PBE functional as implemented in the FHIaims code with preoptimized structures obtained from CASTEP calculations. Reciprocal space has been sampled using a 2 x 2 x 2 Monkhorst-Pack \citep{PhysRevB.13.5188} k-point grid. The structures were fully relaxed until forces have been below 10\textsuperscript{-2} eV/\AA \ applying a \textit{light tier1} basis set.
\section{Polaron calculations}
\subsection{Localizing the Polarons}
\label{obtaining_scf_polarons}
From a methodological point of view, a key aspect of DFT+U is the choice of the correct projector-dependent Hubbard U value.\citep{QUA:QUA24521,Kick2019JCTC} A common strategy would be to choose the value according to the experimentally determined defect position within the band gap.\citep{dupuis,Kick2019JCTC} However, owing to the relative novelty of blue LTO, there are no such results to be found in the literature. From TiO\textsubscript{2} the gap state position is experimentally known to be about $1$ eV below the conduction band minimum (CBM)\citep{PhysRevLett.104.036806}. Arguably, titanium atoms in both LTO an TiO\textsubscript{2} are embedded in a similar chemical environment, leading to a similar position of the defect state within the band gap. This assumption is further supported by the observation that both defective TiO\textsubscript{2} and LTO show the same blue color. \cite{Diebold200353,nasara,blue_LTO} Based on this approximation, all our analysis are conducted with a U value of $2.65$ eV, yielding gap states at around 1 eV below the CBM for the most stable polaronic configurations found. 
For localizing the electrons at different Ti atoms we made use of the matrix control approach\citep{PhysRevB.79.235125,matrix_control}. We started with the occupation matrices obtained from a pure PBE\citep{PBE} run and used this as input for the matrix control routines. To obtain electron localization in specific orbitals on a specific Ti atoms, we modified the corresponding diagonal matrix element to 1. Afterwards, full geometry optimization is performed by fixing this modified occupation matrix. To obtain full self-consistency, we used the obtained geometry and the wave function information as input for a second run without constraining the occupation matrix. Table \ref{rel_stabilities} contains the complete list of calculated polaron systems and their relative stabilities. The relative stabilities have been calculated according to
\begin{align}
E_{\rm rel} = E_{\rm tot, i } - E_{\rm tot,\ most\ stable\ structure} \quad,
\end{align}
where $E_{\rm tot, i }$ is the total energy of a specific configuration and $E_{\rm tot,\ most\ stable\ structure}$ is  the total energy of the most stable configuration. By this, all other less stable configurations show positive energies.\citep{dupuis} 

\begin{table}
\caption{Relative stabilities of the calculated polaronic configurations. Here, we only considered triplet configurations. Furthermore,  d$_{\rm V_O}$ indicates the distance form the Ti\textsuperscript{3+} center to the next periodic oxygen vacancy in the unit cell. Only the distances within the unrelaxed defect supercell have been considered.}
\label{rel_stabilities}
\begin{tabular}{lccclcc}
\hline\hline
system &  E$_{\rm rel}$ [eV] & d$_{\rm V_O}$[\AA] & & system &  E$_{\rm rel}$ [eV] & d$_{\rm V_O}$[\AA] \\ 
\hline
L3-7/L2-9  & 0.00 & 6.52/7.47 && L3-9/L3-12 & 0.23 & 5.05/6.55 \\
L3-7/L2-11 & 0.01 & 6.52/8.63 && L2-8/L2-12 & 0.23 & 9.45/7.38 \\
L3-7/L2-10 & 0.01 & 6.52/8.50 && L2-8/L2-11 & 0.23 & 9.45/8.63 \\
L3-7/L2-7  & 0.02 & 6.52/7.43 && L3-9/L3-10 & 0.26 & 5.05/4.92\\
L2-7/L3-5  & 0.07 & 7.43/4.95 && L5-1/L3-9 & 0.43 & 2.08/5.05\\
L3-8/L3-4  & 0.12 & 6.51/4.93 && L5-1/L3-12 & 0.45 & 2.08/6.55\\
L3-11/L3-12& 0.12 & 6.34/6.55 && L3-7/L1-3  & 0.49 & 6.52/6.69\\
L3-7/L3-12 & 0.13 & 6.52/6.55 && L4-2/L4-4  & 0.75 & 3.71/3.68\\
L2-7/L2-8  & 0.14 & 7.43/9.45 && L4-3/L4-2  & 0.77 & 6.99/3.71\\
L2-7/L2-12 & 0.15 & 7.43/7.38 && L1-1/L1-4  & 0.77 & 6.84/6.70\\
L3-5/L3-6  & 0.19 & 4.95/4.81 && L5-5/L4-2  & 0.82 & 6.22/3.71\\
L3-5/L4-2  & 0.19 & 4.95/3.71 && L4-3/L4-4  & 1.00 & 6.99/3.68\\
L3-7/L3-9  & 0.20 & 6.52/5.05 && L5-1/L5-5  & 2.59 & 2.08/6.22\\
\hline\hline
\end{tabular} 
\end{table}

\begin{figure}
\includegraphics[scale=.4]{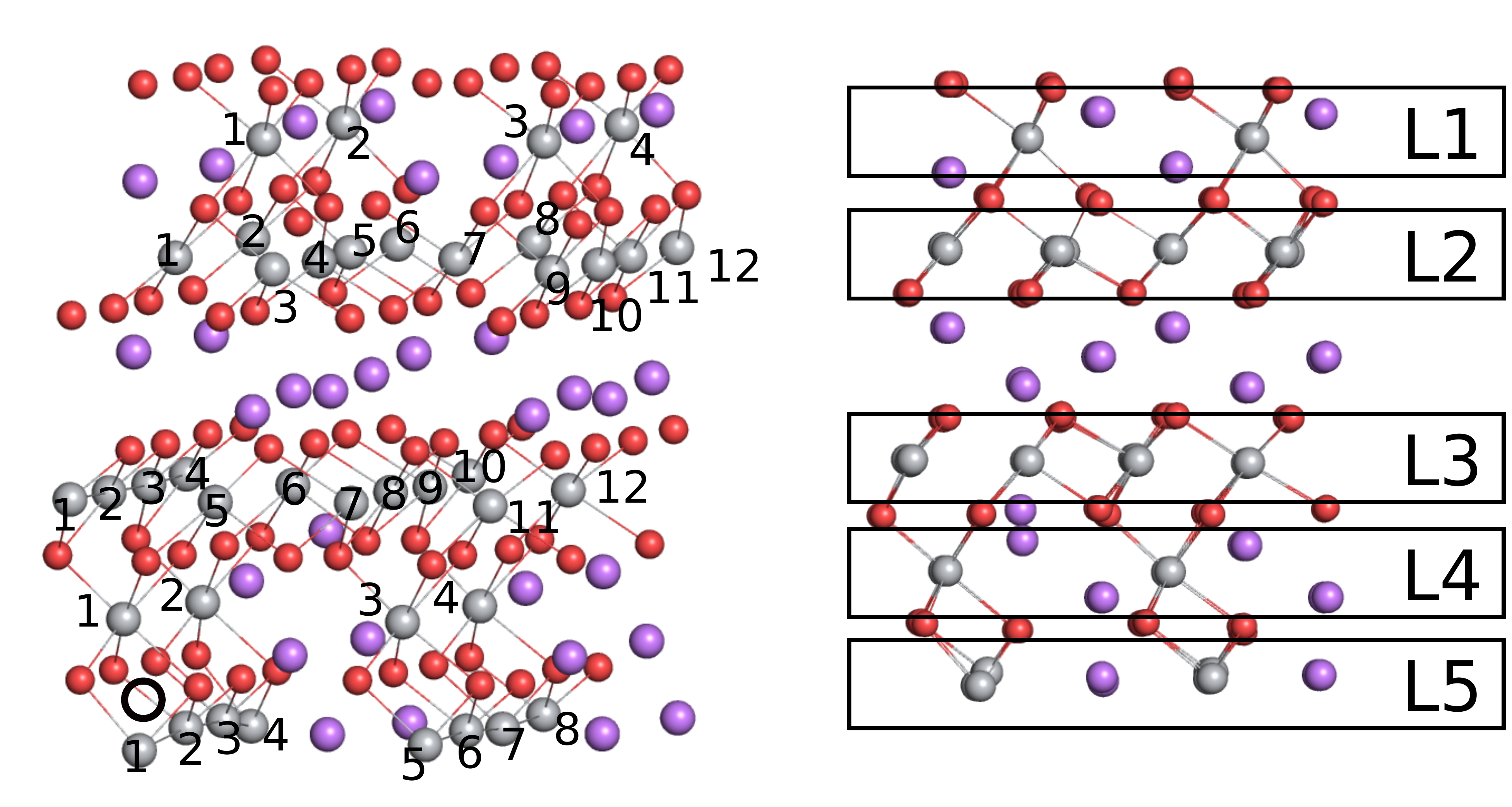}
\caption{Site naming convention of the different localization patterns. L$x$-$m$/L$y$-$n$ thus specifies the localization of one electron within layer L$x$ on atom $m$ with the second electron localized in layer L$y$ on atom $n$. The position of the oxygen vacancy is also marked with a black circle. Titanium atoms are depicted as grey spheres, oxygen atoms are shown as red spheres.}
\label{site_naming}
\end{figure}
%
\subsection{Computational details}
\label{comp_details_polaron_calc}
Electronic structure calculations have been performed entirely using the FHI-aims program package\citep{Blum20092175}.
All structure optimizations have been performed with applying the DFT+U \citep{Hubbard238,anisimov_1,dudarev} variant of the PBE exchange correlation functional using the fully-localized limit (FLL)\citep{QUA:QUA24521} as double counting correction. The Ti $3d$ atomic basis functions serve as Hubbard projectors in each run. Only the on-site representation \citep{ESCHRIG2003482} for the DFT+U occupation matrix have been used in combination with a U value of 2.65 eV. Numerically convergence have been already reached with a \textit{tight tier1} basis set.  Reciprocal space has been sampled using a 2 x 2 x 2 Monkhorst-Pack \citep{PhysRevB.13.5188} k-point grid. The structures were fully relaxed until forces have been below 10\textsuperscript{-2} eV/\AA.

If not otherwise noted, all structures related to the polaron calculations have been fully optimized with PBE+U(=2.65)  only.

\section{Barrier calculations}
\subsection{Obtaining the barrier profile}
\label{obtaining_the_reaction_barrier_profile}

For the barrier profiles we again make use of the matrix control approach. However, instead of fixing all the occupation matrices we only fix the occupation matrices of the atoms between which we assume electron hopping to occur. In detail we studied the hopping of one electron from L3-7/L2-3 to L3-7/L1-9 and from L3-7/L2-10 to L3-7/L2-9. For this we define a reaction coordinate x which describes the occupation matrices of both atoms between the hopping process. The occupation matrix $\mathbf{n}'$ for a certain atom A and a certain atom B is then defined according to
\begin{subequations}
\begin{align}
\label{reaction_coordinate}
\mathbf{n}'_A = x\mathbf{n}^{\rm Ti^{3+}}_A + (1-x)\mathbf{n}^{\rm Ti^{4+}}_A\qquad,\\
\label{reaction_coordinate_II}
\mathbf{n}'_B = (1-x)\mathbf{n}^{\rm Ti^{3+}}_B + x\mathbf{n}^{\rm Ti^{4+}}_B\qquad.
\end{align}
\end{subequations}
$\mathbf{n}^{\rm Ti^{3+}}$ is the occupation matrix where the excess electron is fully localized at a certain atom, this occupation matrix is obtained from a full self-consistent calculation as it is described in section \ref{obtaining_scf_polarons}. $\mathbf{n}^{\rm Ti^{4+}}$ is the occupation matrix at a certain atom if none of the excess electrons is localized. Going from L3-7/L2-10 to L3-7/L2-9  this means $\mathbf{n}^{\rm Ti^{3+}}_A$ belongs to the self-consistent occupation matrix of L2-9 in system L3-7/L2-9 and $\mathbf{n}^{\rm Ti^{4+}}_A$ belongs to the occupation matrix of atom L2-9 in the L3-7/L2-10 system. Occupation matrix $\mathbf{n}^{\rm Ti^{3+}}_B$ would then be the self-consistent occupation matrix of atom L2-10 in system L3-7/L2-10 and $\mathbf{n}^{\rm Ti^{4+}}_B$ corresponds to the self-consistent occupation matrix of atom L2-10 in system L3-7/L2-9. As a next step we performed full structure relaxations with applying the above described constraints. This means that our chosen reaction coordinate directly translates also to a structural change along the reaction path. Table \ref{barriers} lists the obtained energies with respect to the configuration lowest lying in energy.
\begin{table}
\caption{Listed are the points calculated for the barrier profile which is shown in figure \ref{barrier_profile}. All points have been calculated from the matrix-control run. The lowest energy configuration serves as zero point for the energy scale. The notation in brackets corresponds to the self-consistent polaron configuration.}
\label{barriers}
\begin{tabular}{c c c}
\hline \hline
x & L3-7/L2-10 $\rightarrow$ L3-7/L2-9 & L3-7/L2-3 $\rightarrow$ L3-7/L1-9 \\ 
  & [eV] & [eV] \\
  \hline
0.00  & 0.008 & 0.485  \\
  &(L3-7/L2-10) & (L3-7/L1-3)\\
0.10 & 0.024 & 0.501 \\
0.25 & - & 0.565 \\
0.30 & 0.123 & 0.583 \\
0.40 & 0.170 & - \\
0.45 & - & 0.483 \\
0.50 & 0.186 & 0.416 \\
0.60 & 0.166 & 0.280 \\
0.70 & 0.118 & 0.162 \\
0.80 & -   & 0.074 \\
0.90 & 0.016 & 0.019 \\
1.00 & 0.000 & 0.000  \\
 & (L3-7/L2-9) & (L3-7/L2-9) \\ 
\hline \hline
\end{tabular} 
\end{table}
\begin{figure}
\includegraphics[scale=.6]{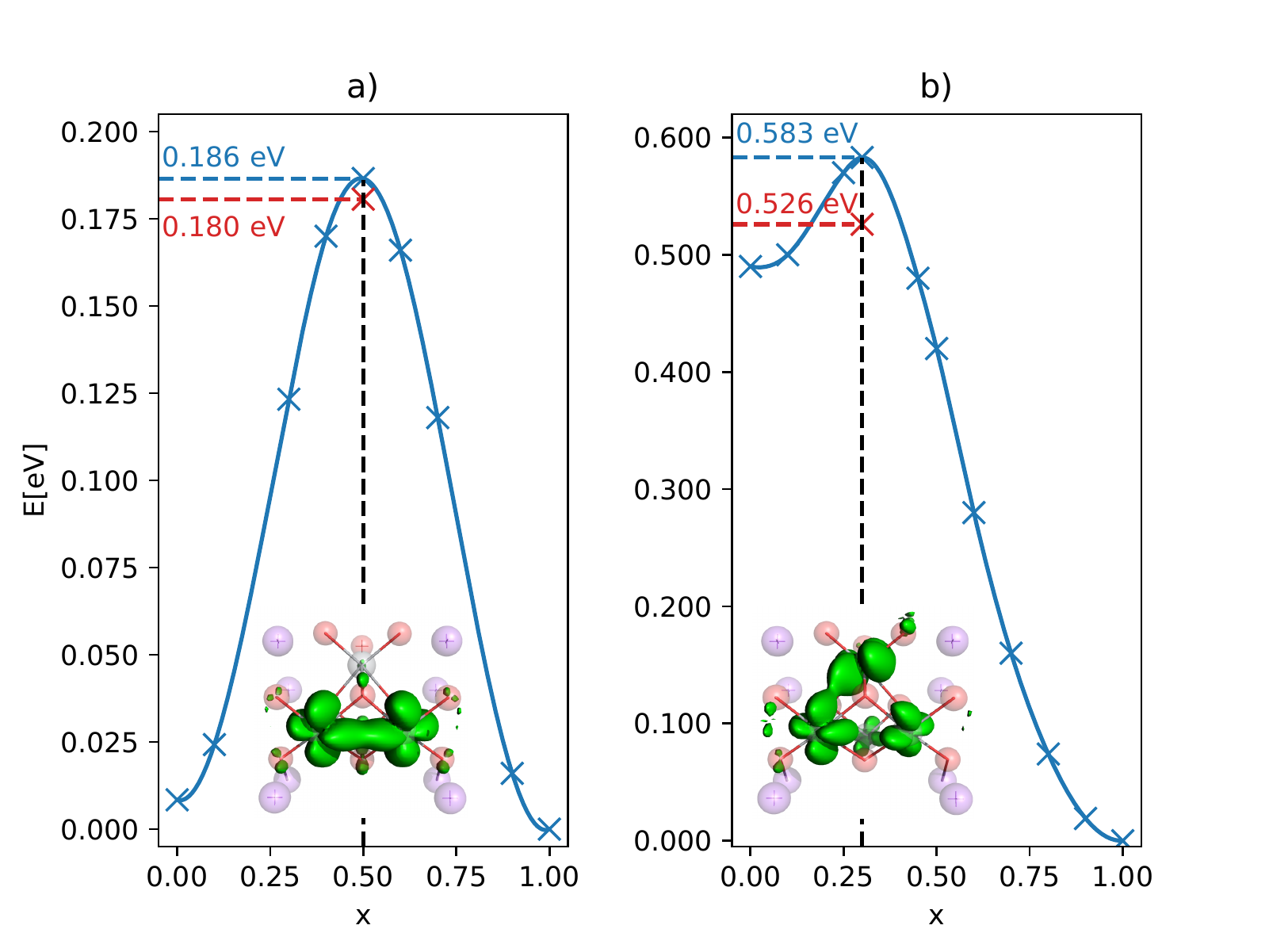}
\caption{Calculated barrier profile for the transition L3-7/L2-10 to L3-7/L2-9 (a) and L3-7/L1-3 to L3-7/L2-9 (b). x = 1.0 is equal to configuration L3-7/L2-9. Blue points are energies obtained with fixed occupation matrix according to eq. \ref{reaction_coordinate} in both subfigures. Red points are DFT+U energies calculated from the wave-function obtained from the matrix control run. The lowest lying configuration in energy serves as zero point for the energy scale. Also shown is the spin density of the corresponding transition state. Isosurface level 0.015 e$\rm\AA^{-3}$.}
\label{barrier_profile}
\end{figure}
However, this is the energy with applied constraints from the matrix control approach. In addition, in order to further judge the quality of this approach we use the obtained
wave-function information from the saddle point and evaluate the DFT+U
functional again without constraining the occupation matrices (red points in
Figure \ref{barrier_profile}). This gives a direct estimate how the electron
density is able to adapt to the applied bias potential. For the transition
L3-7/L2-10 to L3-7/L2-9 depicted in Figure \ref{barrier_profile}a both
energies agree remarkable well, with a difference of only about $6$ meV. For
the other transition considered here, L3-7/L1-3 to L3-7/L2-9, (Figure
\ref{barrier_profile}b) the difference is with $56$ meV significantly larger.
To analyse this further, we calculated the difference in the sum of the
diagonal elements between applied occupation matrix $\mathbf{n}'$ and the
actual occupation matrix $\mathbf{n}$  determined from the wave function at
the transition state points, $\Delta = \sum _n |n'_{nn}| - |n_{nn}|$. For
transition L3-7/L2-10 to L3-7/L2-9 the highest deviation is for the occupation
of L2-9 with a value of $0.036$. Contrary, for transition
L3-7/L1-3 to L3-7/L2-9 our analysis yields a deviation of $0.176$ for the occupation matrix
at L2-8. This large difference is due to the electron not only being
located at L2-8 and L1-3 but also at neighboring titanium atoms. This might
indicate that the applied bias potential does not fit the transition state
equally well as in the case of the L3-7/L2-10 to L3-7/L2-9 transition and
hence is responsible for the observed difference in barrier height. However,
this problem is closely related to the problem of defining proper fragments in
CDFT.\citep{Oberhofer2010JCP}
Yet, given the fact that this specific transition already shows a significant
energy difference of 485 meV between the two stable states, this discrepancy is
most likely of minor importance. Our new approach is thus capable of
gauging hopping barriers at nearly no overhead over standard DFT calculations.
 Our results which we obtain for a U value of 2.65 eV are listed in table \ref{barriers_II}.
\begin{table}
\caption{Barrier heights calculated from the density obtained from the matrix control run.}
\label{barriers_II}
\begin{tabular}{c c c}
\hline \hline
x & L3-7/L2-10 $\rightarrow$ L3-7/L2-9 & L3-7/L2-3 $\rightarrow$ L3-7/L1-9 \\ 
  & [eV] & [eV] \\
\hline 

0.30 & - & 0.526 \\

0.50 & 0.180 & - \\

\hline \hline
\end{tabular} 
\end{table}
It should be highlighted, that the obtained barriers sensitively depend on the applied U value, however, this is a general aspect of DFT+U and not a result of the here applied strategy for obtaining the barrier profiles. Moreover, hybrid functionals should suffer from this drawback too. The U value in DFT+U determines the strength of the on-site coulomb repulsion and hence it determines the amount of how DFT+U will accounts for the self-interaction error. In hybrid functional the mixing factor determines the amount of exact exchange and hence how much a specific hybrid functional accounts for the self-interaction error. 
\subsubsection{Marcus reaction coordinate}
\begin{figure}
\includegraphics[scale=.7]{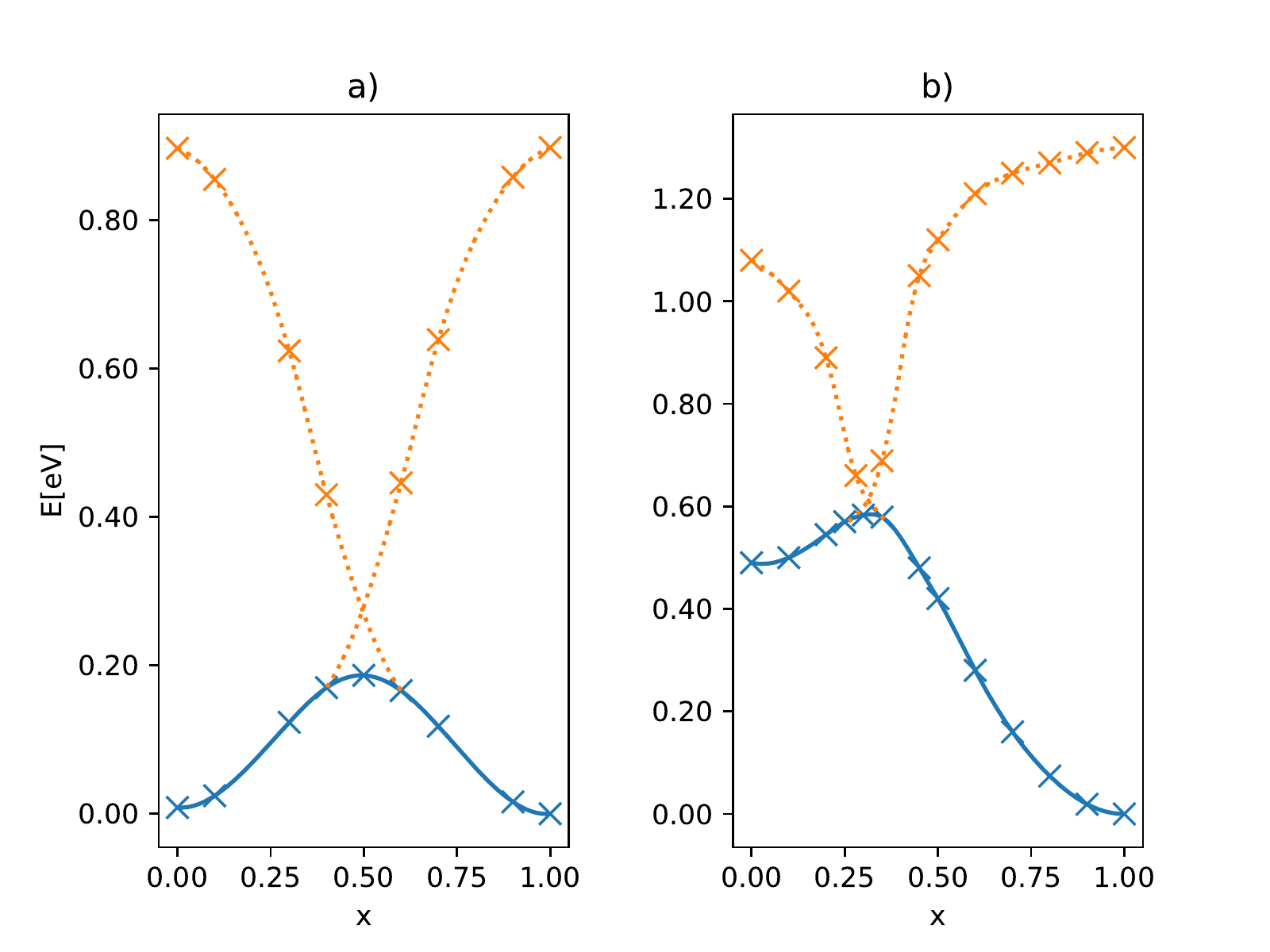}
\caption{Full reaction profile according to reaction coordinate x which defines the occupation matrices according to eq. \ref{reaction_coordinate} and eq. \ref{reaction_coordinate_II} respectively. Blue curve and blue points have been obtained according to the description in section  \ref{obtaining_the_reaction_barrier_profile}.  L3-7/L2-10 to L3-7/L2-9 is shown in (a) and L3-7/L1-3 to L3-7/L2-9  is depicted in (b). x = 1.0 is equal to configuration L3-7/L2-9.}
\label{marcus_reaction_coordinate_I}
\end{figure}

Going along reaction coordinate x the polaron hopping from atom A to atom B occurs as the system passes the transition state. By again exploiting the matrix control approach one is able to constrain the electron at a certain atom and thus preventing the electron to hop. In figure \ref{marcus_reaction_coordinate_I} we are showing the corresponding results. Blue dots have been obtained as described in section \ref{obtaining_the_reaction_barrier_profile}, for orange dots we use the structure obtained for a certain x and use the occupation matrix control to constrain the polaron to one specific site. No geometry relaxation have been allowed during that procedure. 

\begin{figure}
\includegraphics[scale=.7]{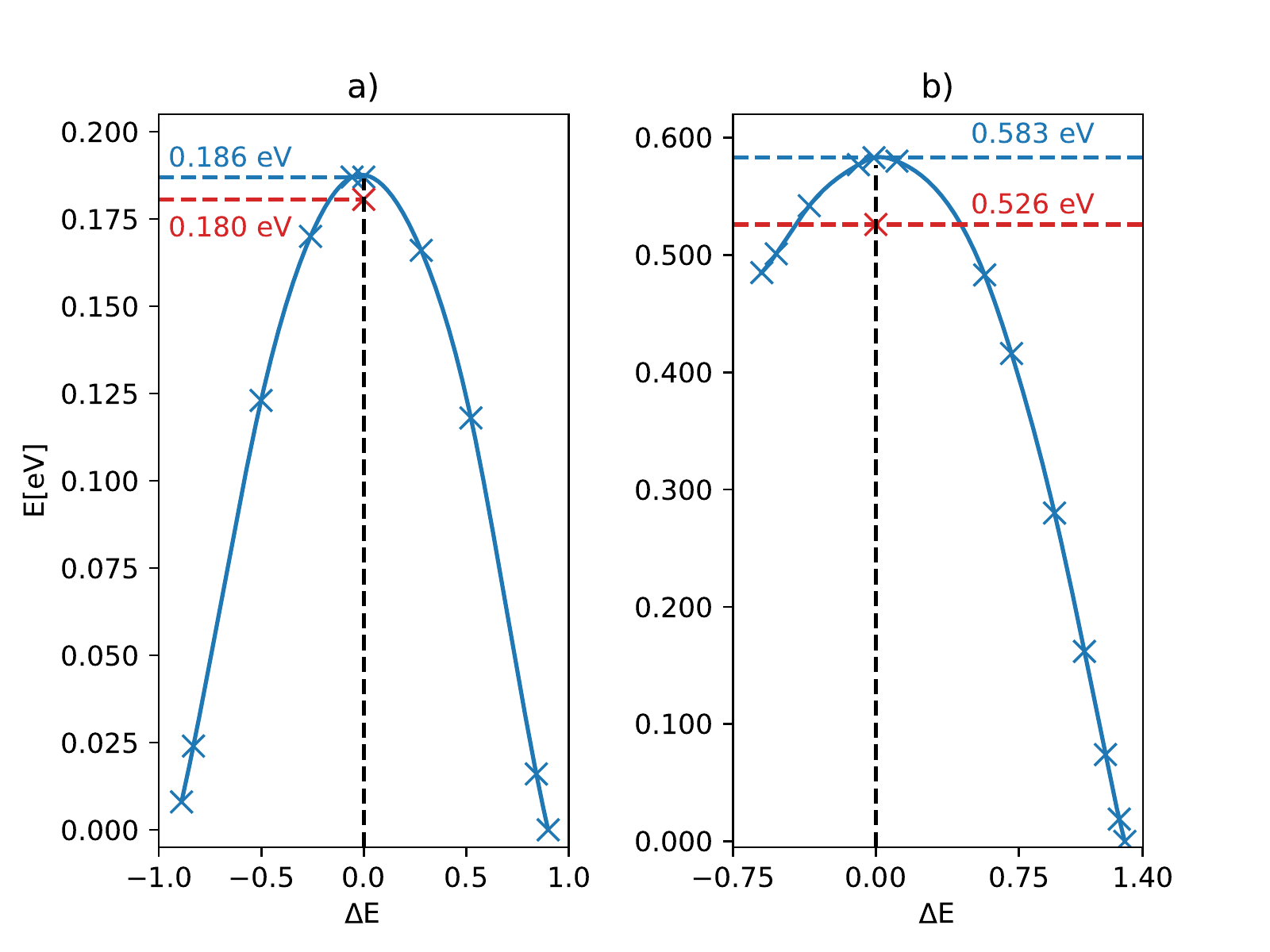}
\caption{}
\label{marcus_reaction_coordinate_II}
\end{figure}

In a next step, this allows to define a new reaction coordinate according to, 
\begin{align}
\Delta E = E\left(x,A\right)-E\left(x,B\right)\quad.
\end{align}
Here, $E\left(x,A\right)$ is the energy if the electron is located on atom A. Whereas $E\left(x,B\right)$ is the energy if the electron is located at atom B. For both energies the same geometry has been used. In Figure \ref{marcus_reaction_coordinate_II} we show the new reaction profile. Clearly, Figure~\ref{marcus_reaction_coordinate_II}b indicates an early transition state for the hopping from L3-7/L1-3 to L3-7/L2-9.
\subsection{Computational details}
For the barrier calculations we applied the same computational settings as described in section \ref{comp_details_polaron_calc}.
\subsection{Conductivity model}
Using the barriers $\Delta E$ of 186\,meV and 583\,meV for the in- and out-of-plane polaron transfer, respectively, 
as representative for the hopping in the LTO crystal, we first compute the respective hopping rates using
a simple transition state theory:\cite{Oberhofer2017CR}
\begin{equation}
	k_\text{hop}=\frac{k_\text{B}T}{h}e^{-\Delta E/k_\text{B} T}
\end{equation}
Here, $k_\text{B}$, $T$, and $h$ denote the Boltzmann constant, the temperature, and Planck's constant, respectively.
As a first estimate we here also assume entropic effects to play a minor role for the electronic transition considered
here. 

These rate constants can then be used in a simple Einstein-Smoluchowsky model of the carrier mobility\cite{Oberhofer2017CR}
\begin{equation}
 \mu=\frac{e a^2 k_\text{hop}}{k_\text{B} T},
\end{equation}
where $e$ is the elementary charge and $a$ is the distance between the initial and final polaron localization sites.
The such calculated mobilities finally allow us to estimate the system's conductivity $\sigma$ using the number 
density of charge carriers $n$ and\cite{Oberhofer2017CR}
\begin{equation}
 \sigma= e n \mu.
\end{equation}

Thereby, $n$ has been taken from experimental measurements of LTO to be $13.1 \text{at}\%$ which suggest $2.9 \times 10^{-3}$ \AA{}$^{-3}$ for our simulation cell. Using a temperature of $300$ K we arrive at conductivities of
$9.53\times 10^{-02}$ S/cm and $1.7\times 10^{-08}$ S/cm for the in- and out-of-plane polaron hopping mechanisms, respectively.

\bibliography{bib.bib}